\documentclass{IEEEcsmag}

\usepackage[colorlinks,urlcolor=blue,linkcolor=blue,citecolor=blue]{hyperref}
\expandafter\def\expandafter\UrlBreaks\expandafter{\UrlBreaks\do\/\do\*\do\-\do\~\do\'\do\"\do\-}
\usepackage{upmath,color}
\usepackage{tcolorbox}
\usepackage{makecell}
\usepackage{multirow, booktabs, url, tabularray}
\usepackage{colortbl}

\usepackage[backend=biber,maxbibnames=7,maxcitenames=5,sorting=none]{biblatex}
\usepackage{rotating}

\jvol{XX}
\jnum{XX}
\paper{8}
\jmonth{Month}
\jname{Publication Name}
\jtitle{Publication Title}
\pubyear{2024}

\begin{document}

\sptitle{Article Type: Special Issue on Creativity in Software Engineering}

\title{On AI-Inspired UI-Design}

\author{Jialiang Wei, Anne-Lise Courbis, Thomas Lambolais, Gérard Dray}
\affil{EuroMov Digital Health in Motion, Univ Montpellier, IMT Mines Ales, France}

\author{Walid Maalej}
\affil{University of Hamburg, Germany}

\markboth{THEME/FEATURE/DEPARTMENT}{THEME/FEATURE/DEPARTMENT}

\begin{abstract}
Graphical User Interface (or simply UI) is a primary mean of interaction between users and their devices. 
In this paper, we discuss three complementary Artificial Intelligence (AI) approaches for triggering the creativity of app designers and inspiring them create better and more diverse UI designs.
First, designers can prompt a Large Language Model (LLM) to directly generate and adjust UIs. 
Second, a Vision-Language Model (VLM) enables designers to effectively search a large screenshot dataset, e.g. from apps published in app stores. 
Third, a Diffusion Model (DM) can be trained to specifically generate UIs as inspirational images.
We present an AI-inspired design process and discuss the implications and limitations of the approaches.
\end{abstract}

\maketitle

The UI of apps plays a pivotal role in the overall user experience, as it translates complex software functions into visually intuitive elements that users can easily understand and navigate. 
In a recent study, Chen et al.~showed that well-designed UI affects user satisfaction and distinguishes an app from its competitors \cite{Chen:HowShouldImprove:2021}. 
A well-designed UI should be functional, user-friendly, and aesthetically pleasing, enabling users to accomplish their tasks efficiently while leaving them satisfied or even impressed \cite{Chen:HowShouldImprove:2021, Pourasad:ICSE:2024}. 
A good design usually requires creativity \cite{Pham:SEEM:18}, which is the ability to generate novel, diverse and valuable ideas.

The typical app design process involves several key activities as setting design goals, conducting user research, identifying main features and usage scenarios, wireframing, visual design, prototyping, development hand-off, and gathering feedback. 
Commercial tools such as Figma and Sketch provide basic templates and UI components, which serve as valuable starting points for visual design and prototyping. 
However, designers often require several ideation iterations as well additional inspiration to effectively design complex app features and corresponding UIs.
To get inspiration, app designers may, for instance, explore existing UIs on their mobile phones, on app stores such as Google Play and Apple App Store, or on design portfolio platforms like Mobbin\footnote{\url{https://mobbin.com/}} and Dribbble\footnote{\url{https://dribbble.com/}}.

Recent advancement in AI, particularly Generative AI and the availability of large foundation models is revolutionising the process of app development in general and app UI design in particular. 
Recent studies reveal that AI can be effectively used for accurate text-to-UI retrieval \cite{Wei:GUingMobileGUI:2024, Kolthoff:DatadrivenPrototypingNaturallanguagebased:2023}. 
Moreover, Generative AI, encompassing both image and text generation, can be applied to UI generation, offering an immense source of design inspiration \cite{Wei:BoostingGUIPrototyping:2023, Feng:DesigningLanguageWireframing:2023}.
Several commercial tools are already available for text-to-UI generation, such as Uizard's Autodesigner\footnote{\url{https://uizard.io/autodesigner/}}, Galileo AI\footnote{\url{https://www.usegalileo.ai/explore}}, Visily's AI\footnote{\url{https://www.visily.ai/ai-ui-design-generator/}}, or JS.Design's AI\footnote{\url{https://jsai.cc/ai/create}}.
App development teams are starting to explore these trends with one central question: how can AI assist UI design or even automate it?

\begin{figure}
\centerline{\includegraphics[width=0.5\textwidth]{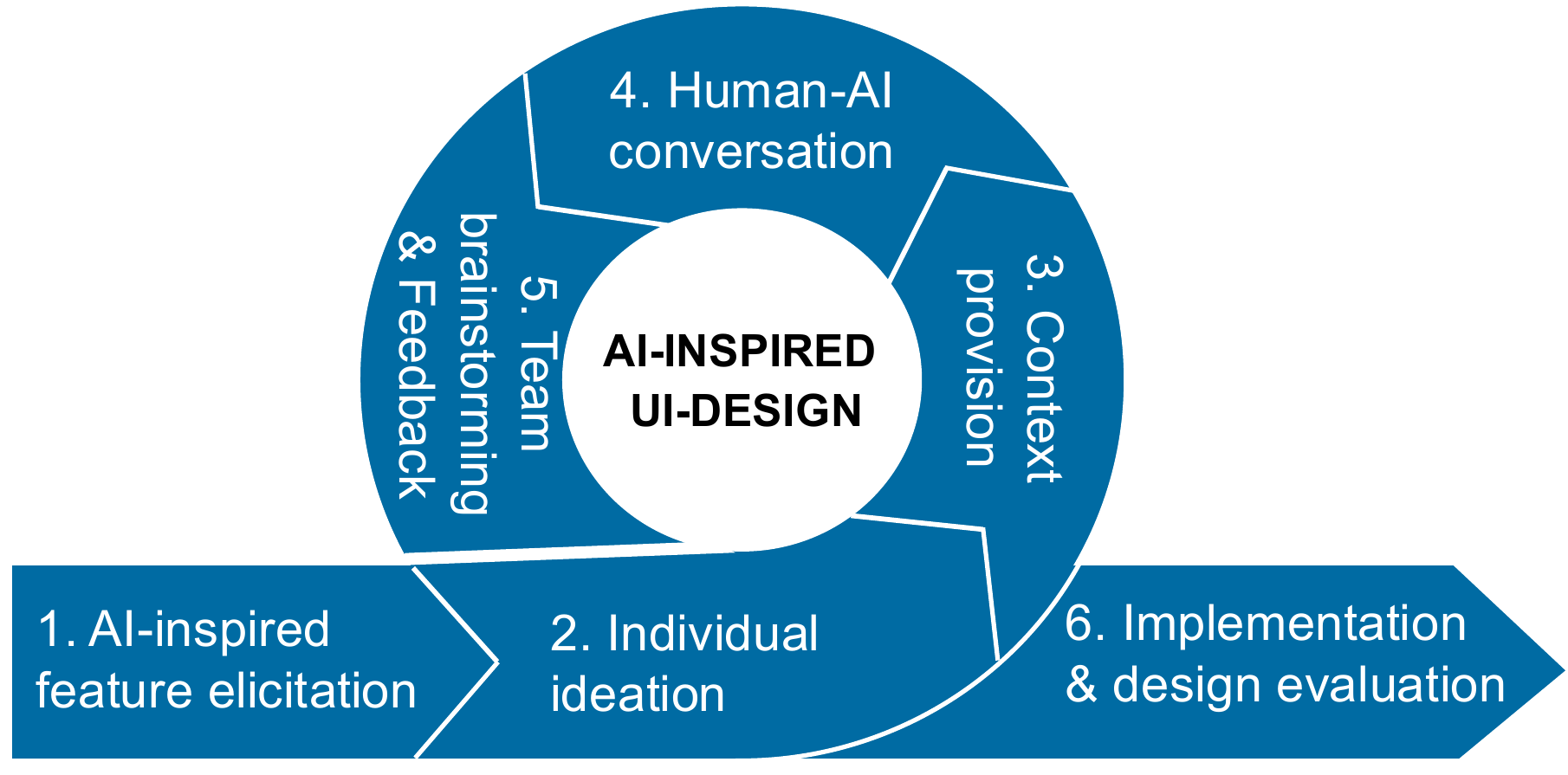}}
\caption{An AI-inspired design process for UI-intensive software (inspired by \cite{Gohar:TR:2024, HBR:2024}).}
\label{fig:aid-process}
\end{figure}

\section{Boosting Creativity with AI}

In a recent fascinating study on creativity, Gohar and Utley investigated the impact of using ChatGPT on creative problem solving and on the quality of generated ideas \cite{Gohar:TR:2024}.
The authors asked practitioner teams from different companies to engage in the creative solving of problems from their organisations. 
They then asked ``product owners'' to assess the quality of resulting ideas (1 to 4 scale from highly compelling to not worth pursuing) while the teams should assess their experiences with the task.
The authors found that teams who used ChatGPT created 8\% more ideas compared to those who did not.  
Interestingly, they observed that Generative AI was \textit{``helping people develop fewer truly bad ideas but AI assistance was also leading to more average ideas''}. 
They concluded that, to outperform in creative problem-solving with AI, teams need to adhere to certain practices while most are not \cite{HBR:2024}. 

What is fascinating about the study is that it shows AI can give \textbf{the impression} to people that they outperform when they simply delegate creative work to AI tools. 
This can be counterproductive as it stimulates human laziness. 
The authors suggested a simple process called FIXIT to guide a AI-supported problem solving, highlighting that AI should be used in a \textbf{conversational} way to get the best of human creativity, instead of a transactional ``do the work for me'' manner. 

In Figure \ref{fig:aid-process} we adopt and extend Gohar and Utley's findings into the context of app design. 
We suggest an AI-inspired app design process, including six steps:

\begin{enumerate}  
    \item First, app designers should scope the requirements as much as possible. 
    Starting from an overall aim or purpose of the app, a list of features, sub-features, and possible user stories can, e.g., be created by prompting an LLM \cite{Wei:GettingInspirationFeature:2024}. 
    
    \item Once a preliminary textual ``backlog'' of features or user stories is available, developers and designers engage in several ideation steps supported by AI tools, first individually and then in team. 
    Research has repeatedly shown the importance of balancing between individual ideation and group brainstorming sessions for creative work. 
    Therefore, it is crucial that designers in the second step take time by themselves to briefly think about the target app feature, user interface, and user experience. 
    \item Third, to get the most fitting and inspiring results from foundation models, a prompt or a query should be as detailed and with as much context as possible. 
    This context can, e.g.~be provided to an AI model in form of a problem statement, notes, the elicited features from step 1, or even some UI sketches and existing app screens.
    
    \item The fourth step is the core of AI-inspired design. 
    A designer should engage in an iterative conversation with an AI model, refining the queries in response to the AI output. The underlying assumption is that optimal results requires preliminary critical evaluation by the human expert and multiple adjustments of the queries.
    For instance, when prompting ChatGPT for idea recommendations, designers can immediately follow-up with additional recommendation requests, ask other LLMs to review the recommendations or explain to ChatGPT why certain recommendations are good or bad. 
    This requires careful critical thinking about the generated AI recommendations. 
    Critical questions include what UIs or design elements make most sense and why? 
    What are most interesting and intuitive designs and why?  
    \item The fifth step focuses on the group or team setting, ideally seeking other perspectives such as a customer, a lead user, a domain or a technology expert. 
    This step can follow typical brainstorming or design thinking processes \cite{Pham:SEEM:18} including the assessment and prioritization of the various design options. 
    The team can use AI foundation models 1) to gather additional ideas as if it would be a team member and 2) to gather feedback, e.g.~stimulating certain scenarios and users.
    At the end, the team should evaluate and decide about the design to implement.
    \item Finally, the design and the corresponding logic will be implemented, tested, released to users and ultimately the quality of the design decisions and the corresponding experience evaluated with end users \cite{Pourasad:ICSE:2024}.
    In some cases, AI-generated UI code can also be reused.   
\end{enumerate}

In the following, we present three state-of-the-art approaches to get UI design recommendations from foundation models:   
1) by prompting Large Language Models (LLM);  
2) by searching screenshot repositories using Vision-Language Models (VLM); and 
3) by training a diffusion model (DM) to generate creative app screens.
Readers can try out the approaches in our accompanying online material\footnote{\url{https://github.com/Jl-wei/ai-gen-ui} and \url{https://github.com/Jl-wei/guing}}. 
We use a health monitoring app as illustrative example (see Table \ref{tab:samples}).

\begin{table*}[]
\setlength{\fboxrule}{1pt}
\setlength{\fboxsep}{0pt}

\caption{Examples of UIs obtained using three different approaches with the input ``Health Monitoring Report''.}
\centering
\begin{tblr}{h{1.5cm}|Q[m,c]|Q[m,c]|Q[m,c]|Q[m,c]|Q[m,c]|Q[m,c]}
\hline[1pt]
\SetCell[r=1,c=1]{m,c}\textbf{Approach} & \SetCell[r=1,c=6]{m,c} {\textbf{UI Examples}}\\ 
\hline
UI Generation with LLMs
& 
\fbox{\includegraphics[height=3.5cm,keepaspectratio]{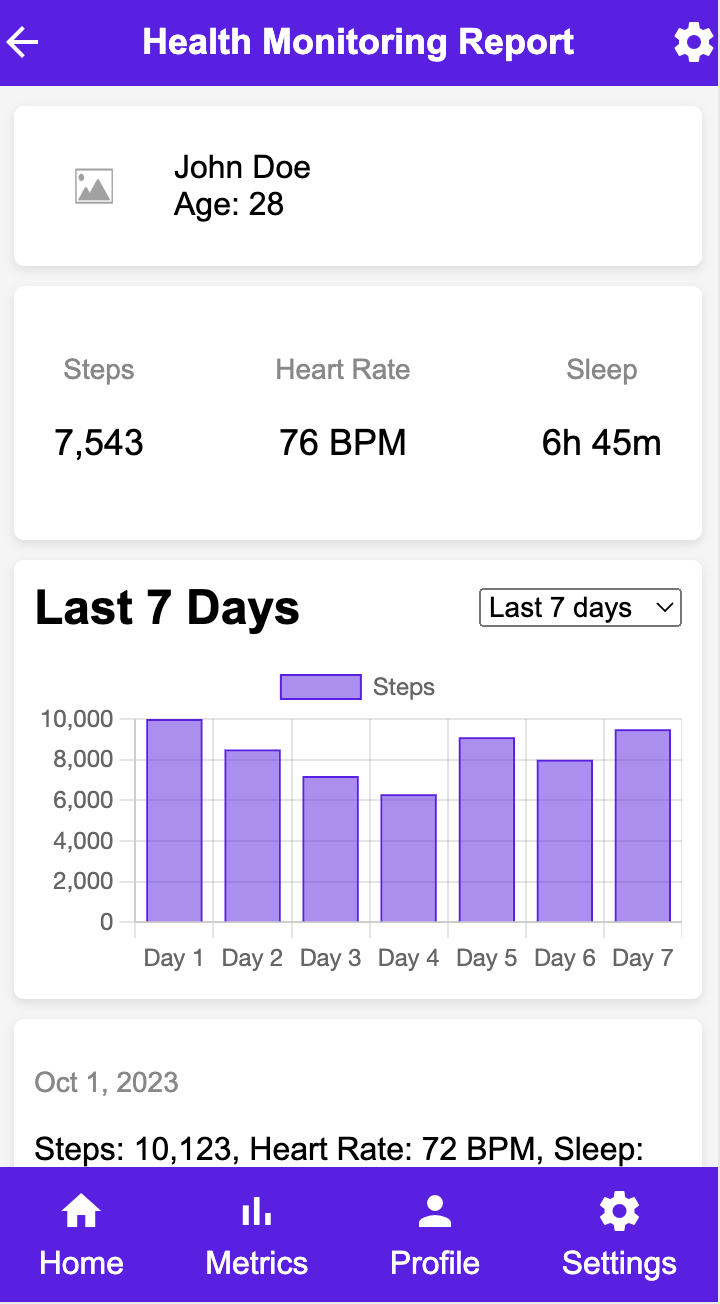}} & 
\fbox{\includegraphics[height=3.5cm,keepaspectratio]{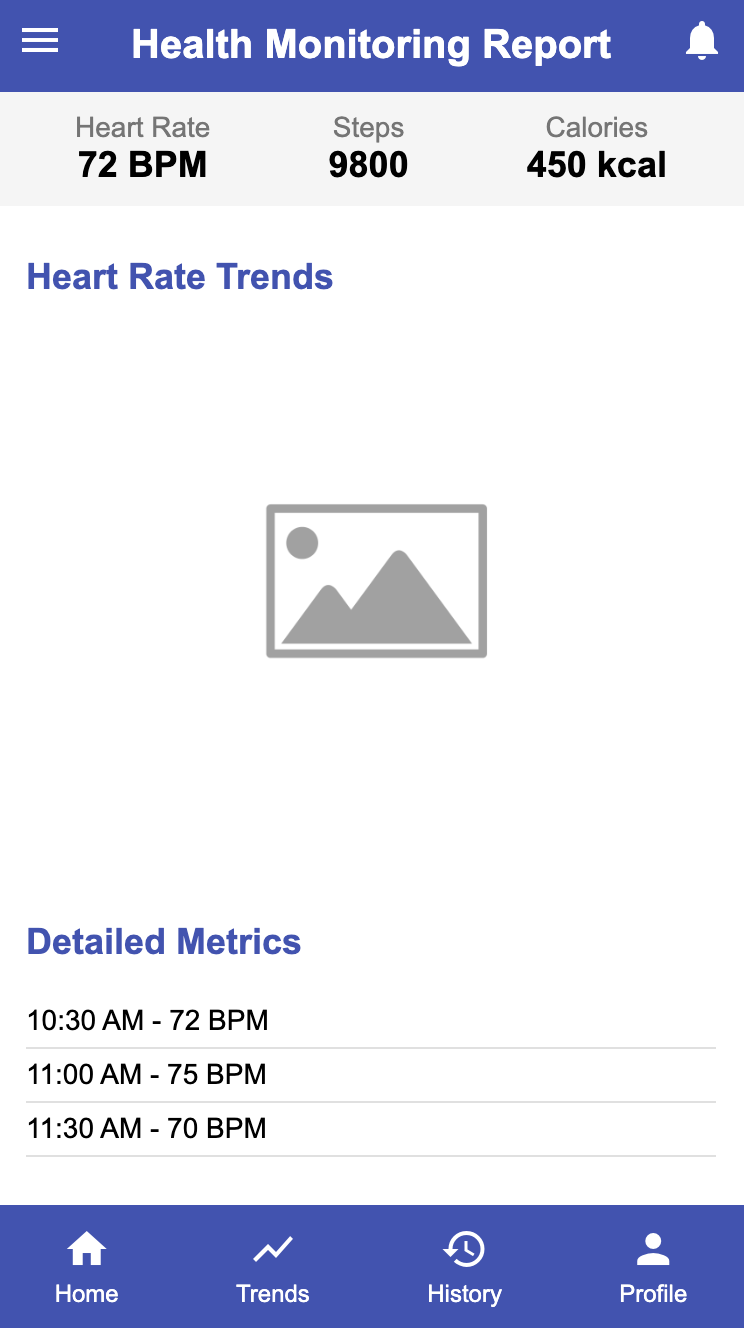}} & 
\fbox{\includegraphics[height=3.5cm,keepaspectratio]{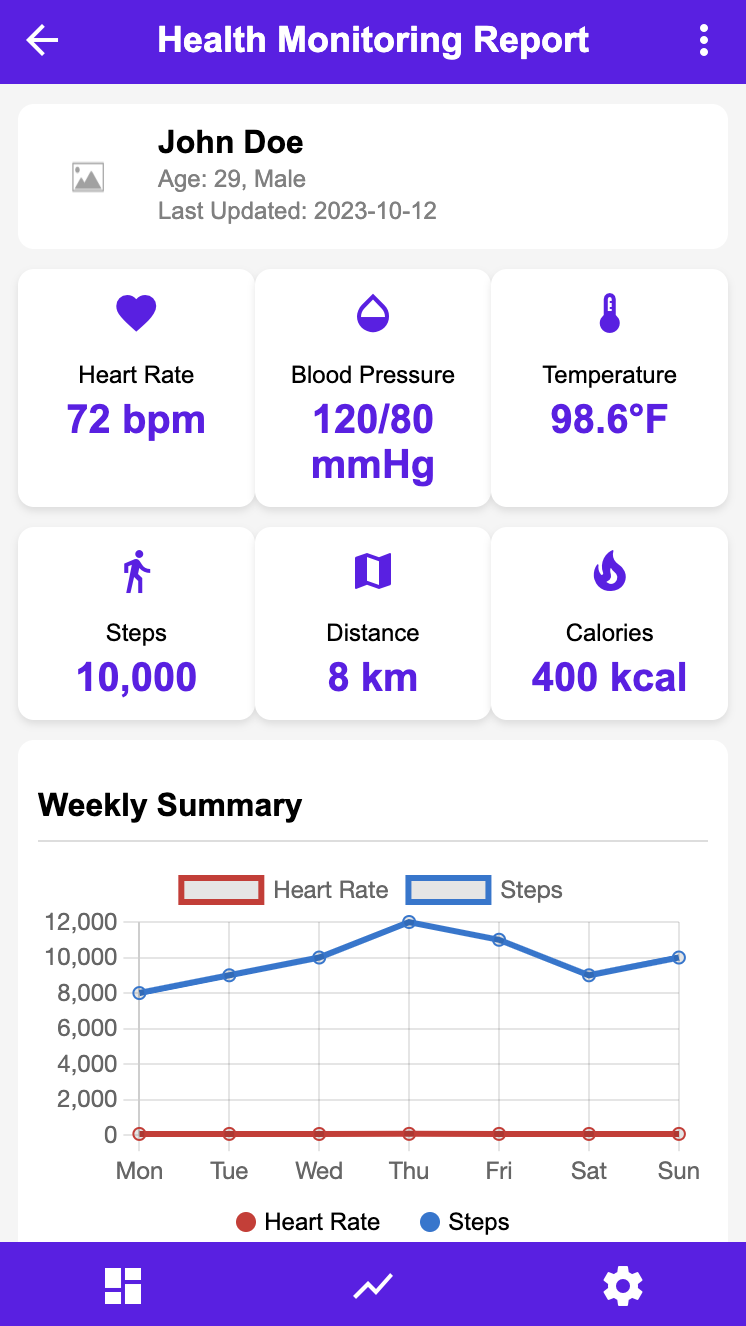}} & 
\fbox{\includegraphics[height=3.5cm,keepaspectratio]{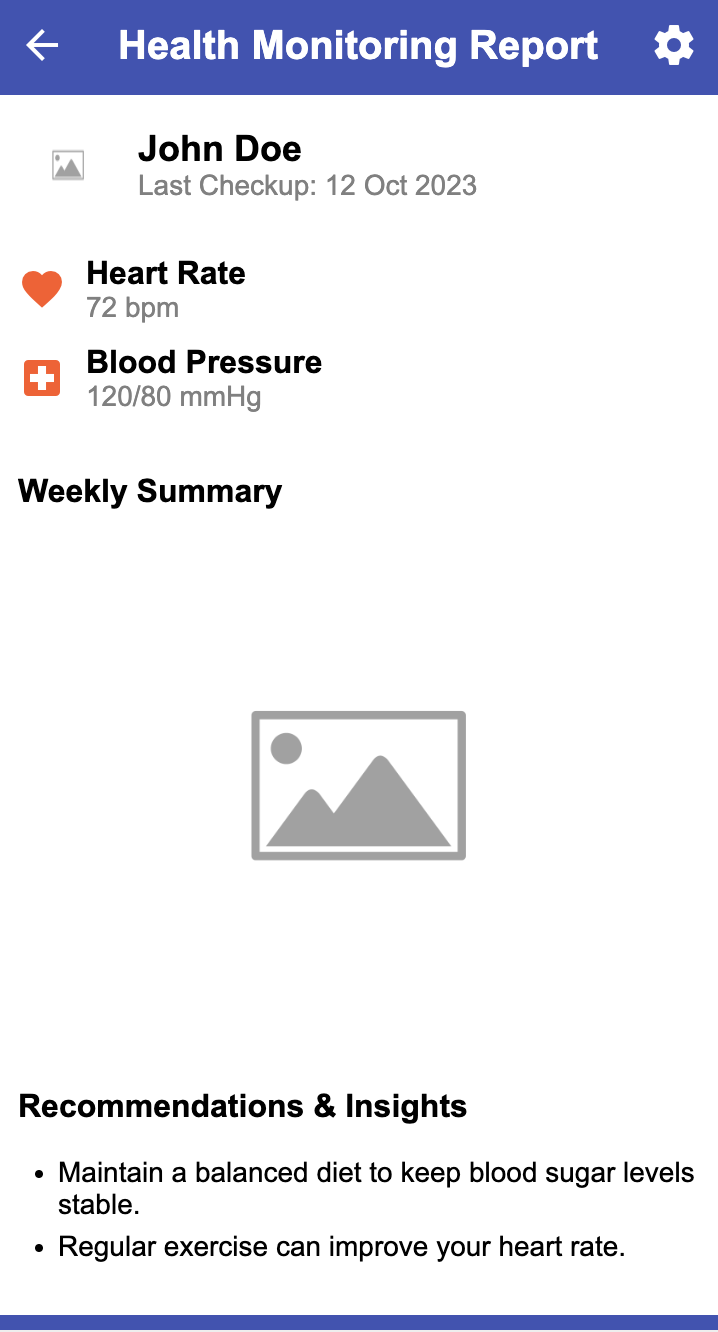}} &
\fbox{\includegraphics[height=3.5cm,keepaspectratio]{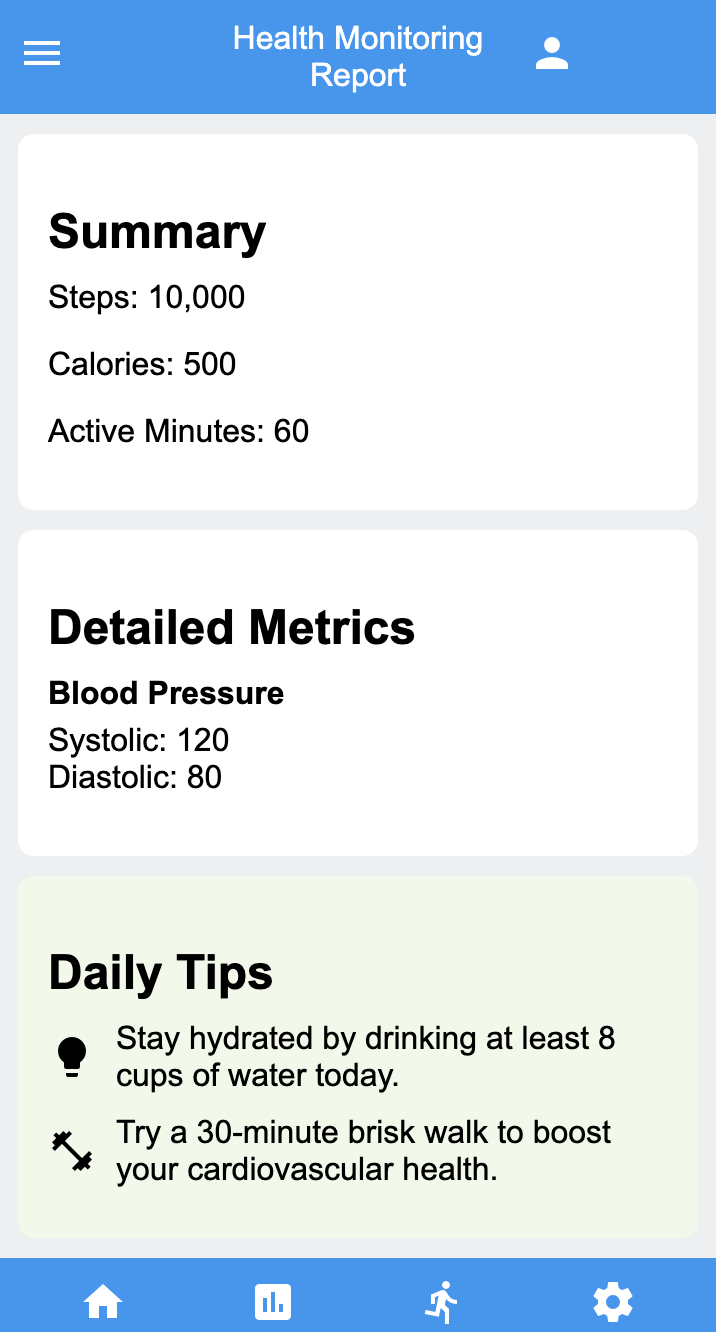}} &
\fbox{\includegraphics[height=3.5cm,keepaspectratio]{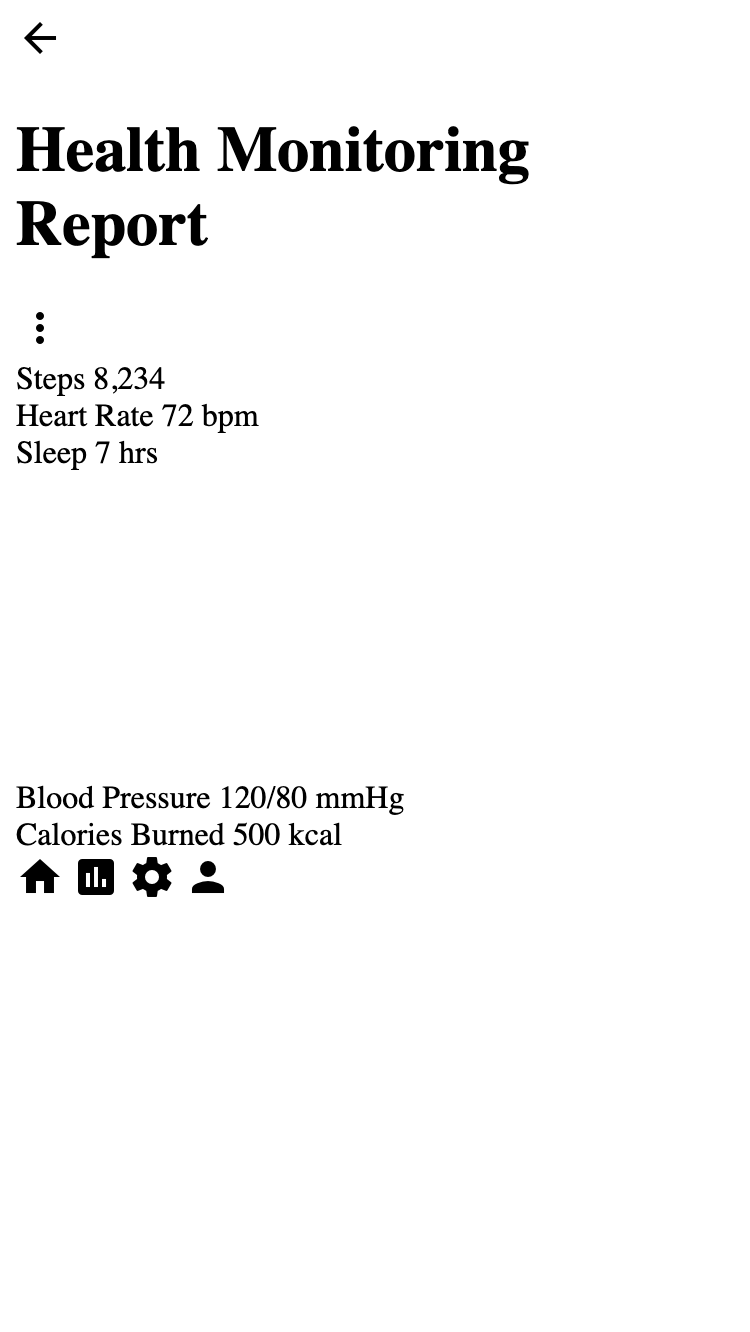}}\\
\hline
UI Retrieval with VLMs
&
\fbox{\includegraphics[height=3.5cm,keepaspectratio]{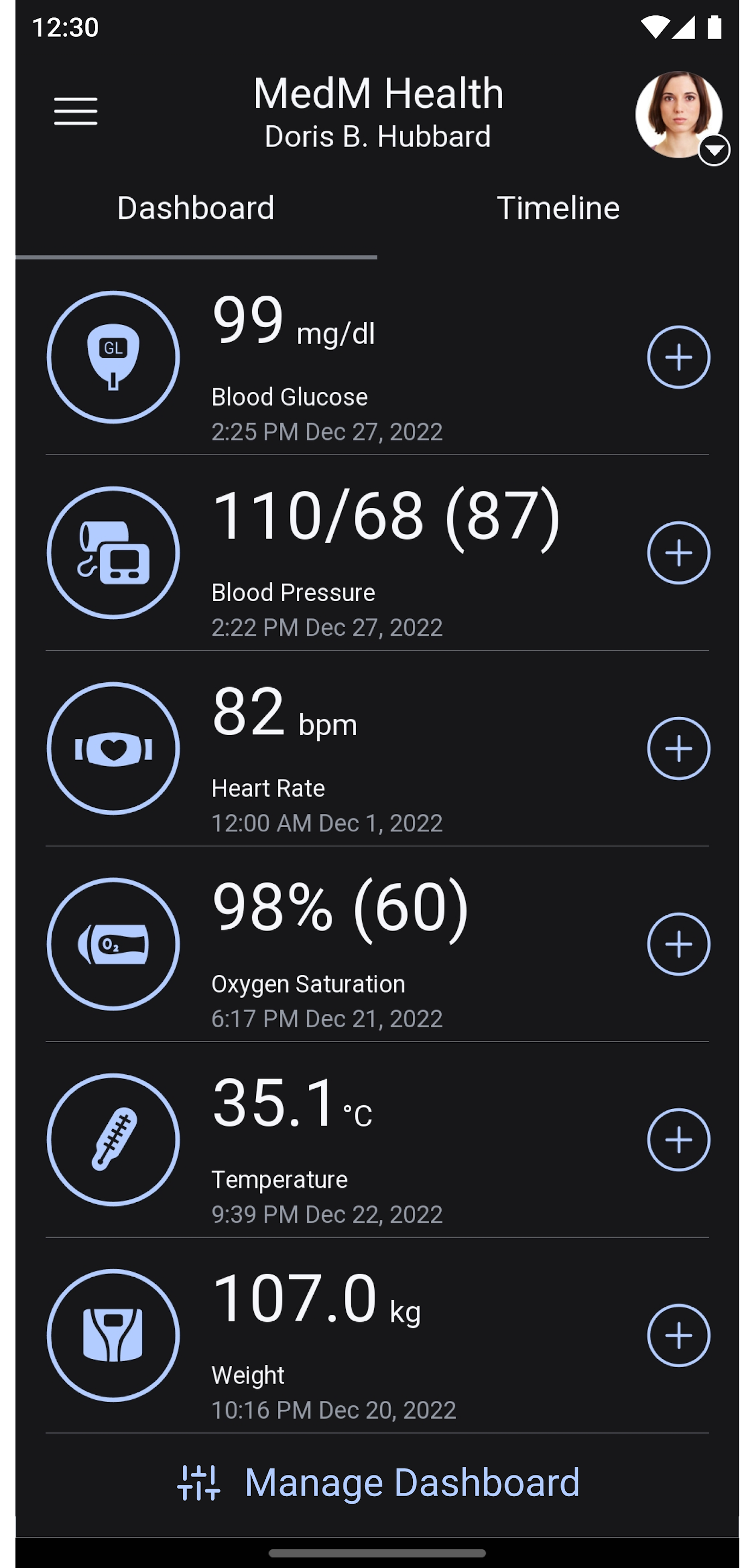}} & 
\fbox{\includegraphics[height=3.5cm,keepaspectratio]{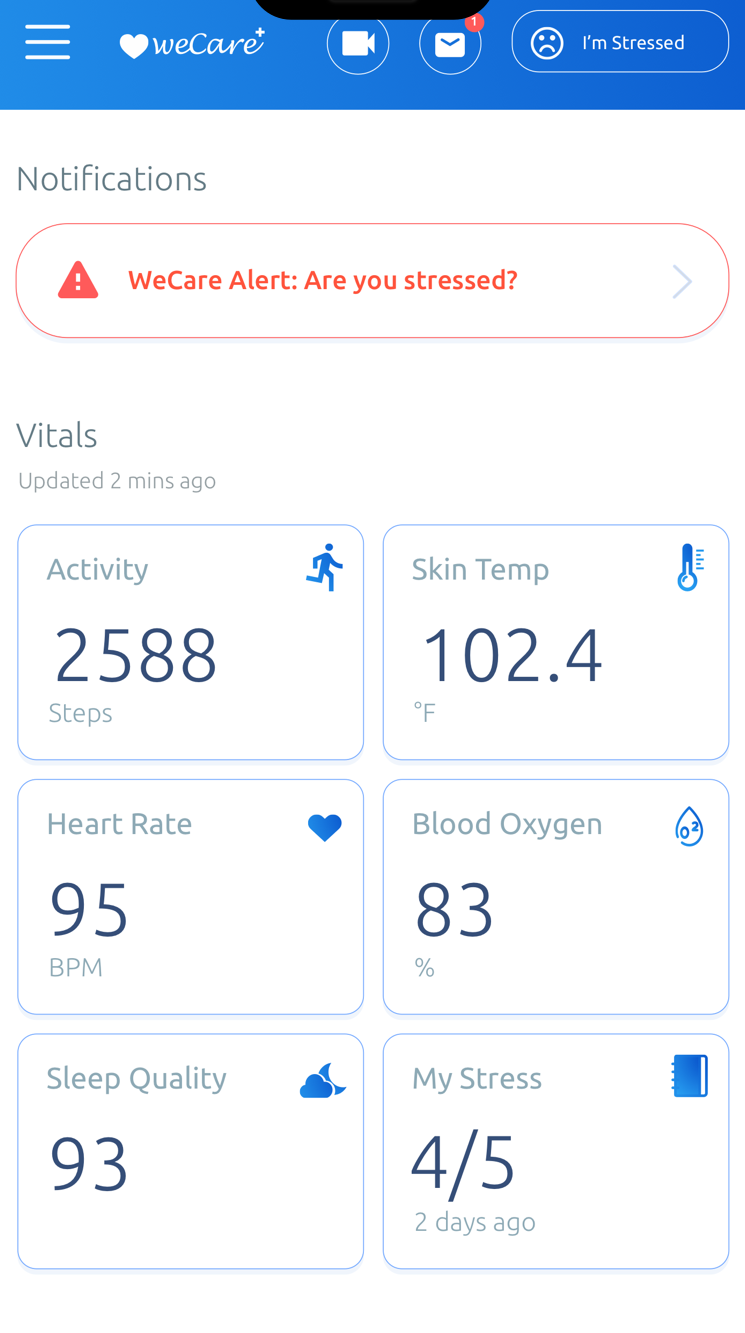}} & 
\fbox{\includegraphics[height=3.5cm,keepaspectratio]{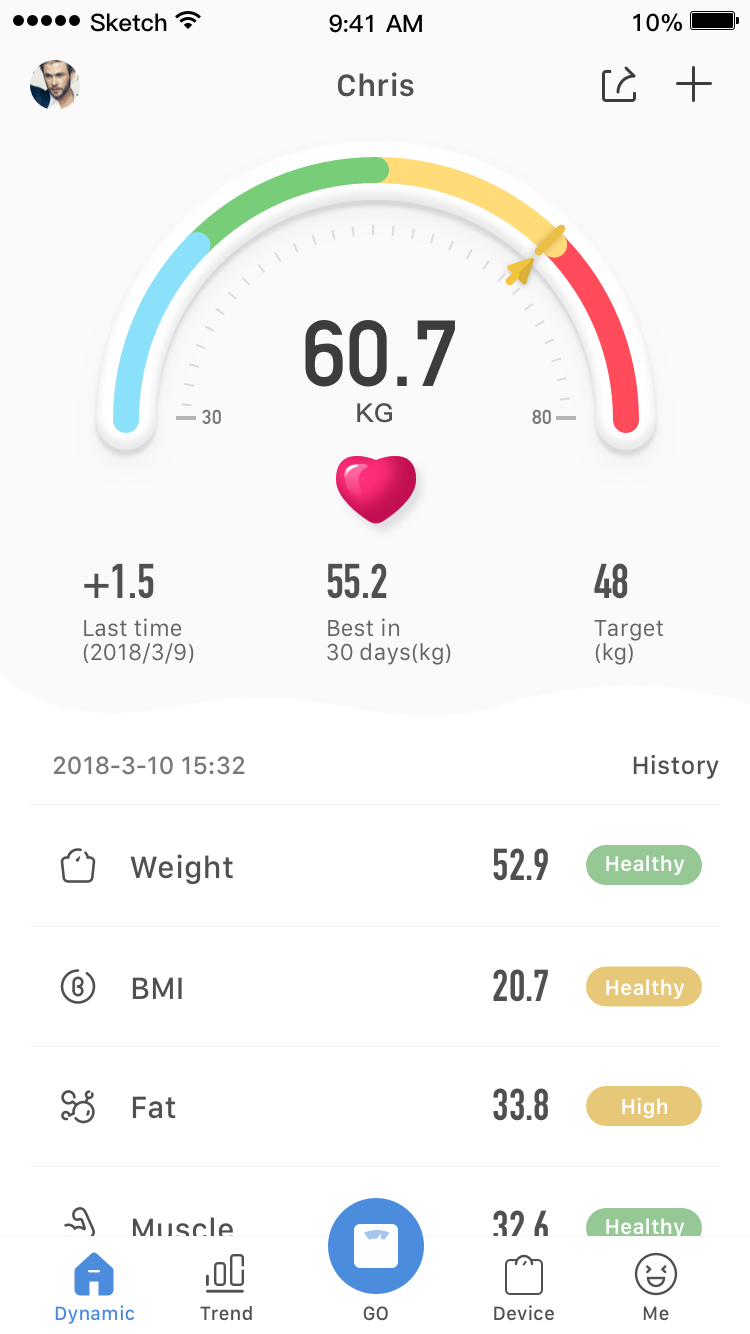}} &
\fbox{\includegraphics[height=3.5cm,keepaspectratio]{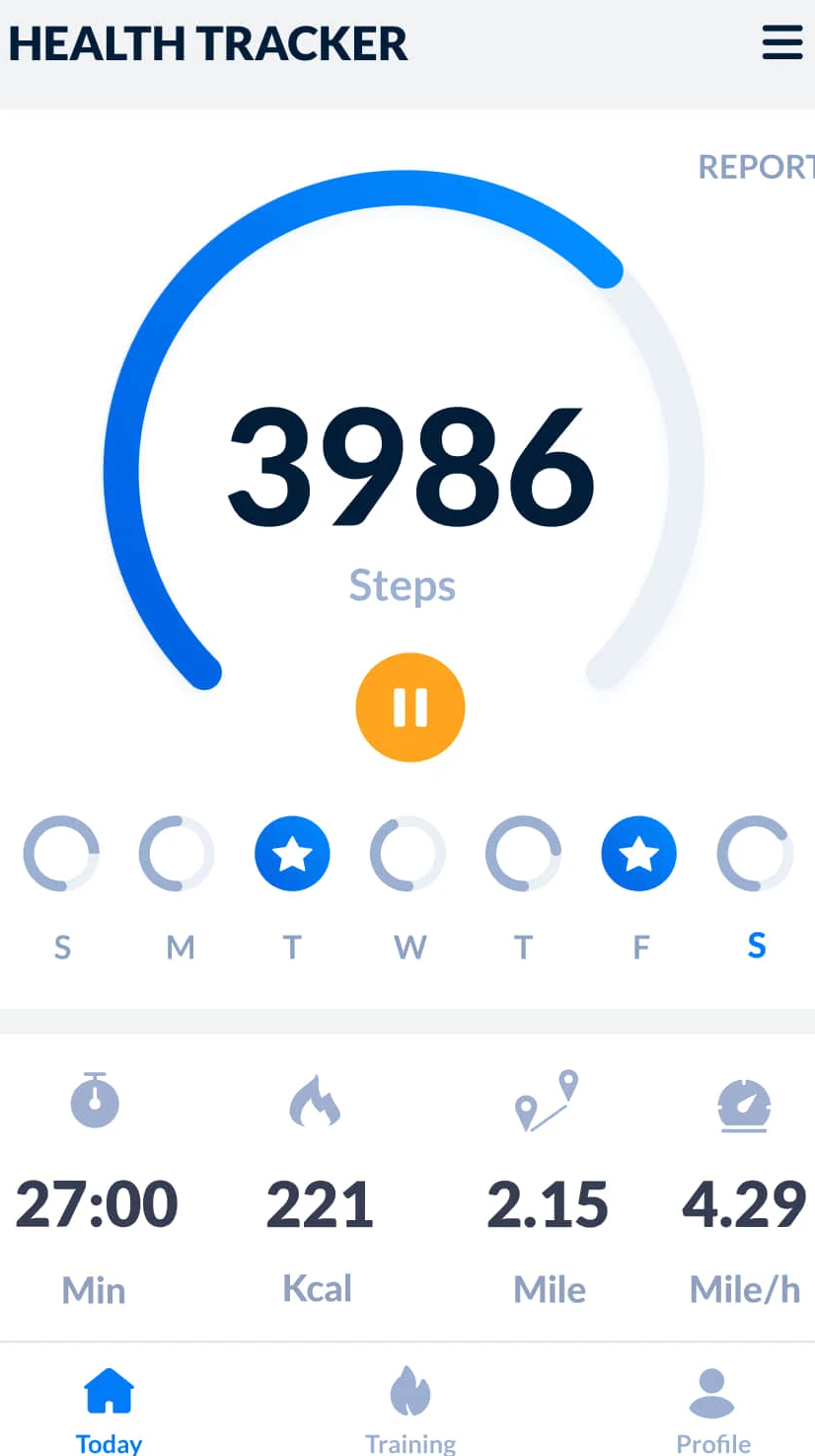}} &
\fbox{\includegraphics[height=3.5cm,keepaspectratio]{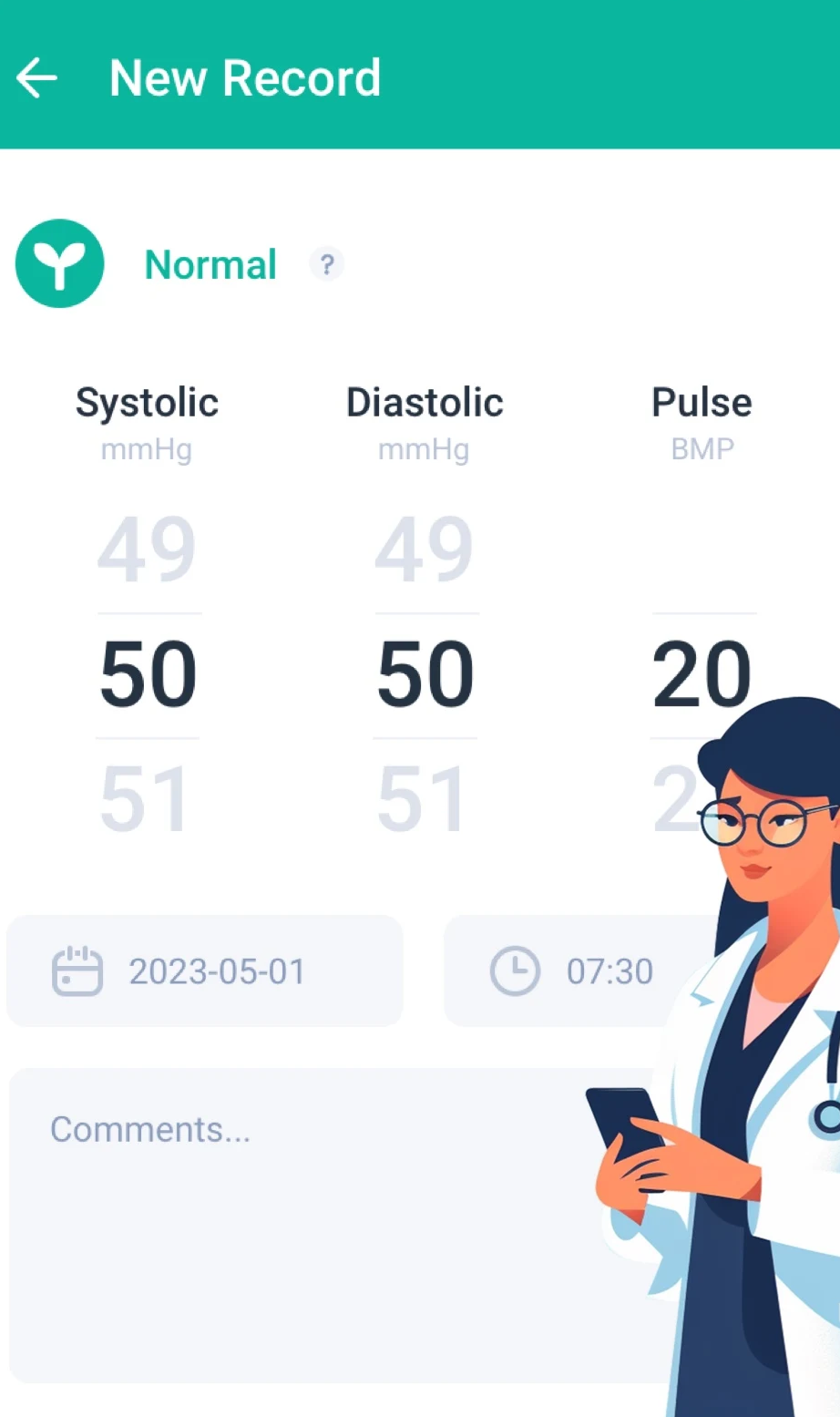}} & 
\fbox{\includegraphics[height=3.5cm,keepaspectratio]{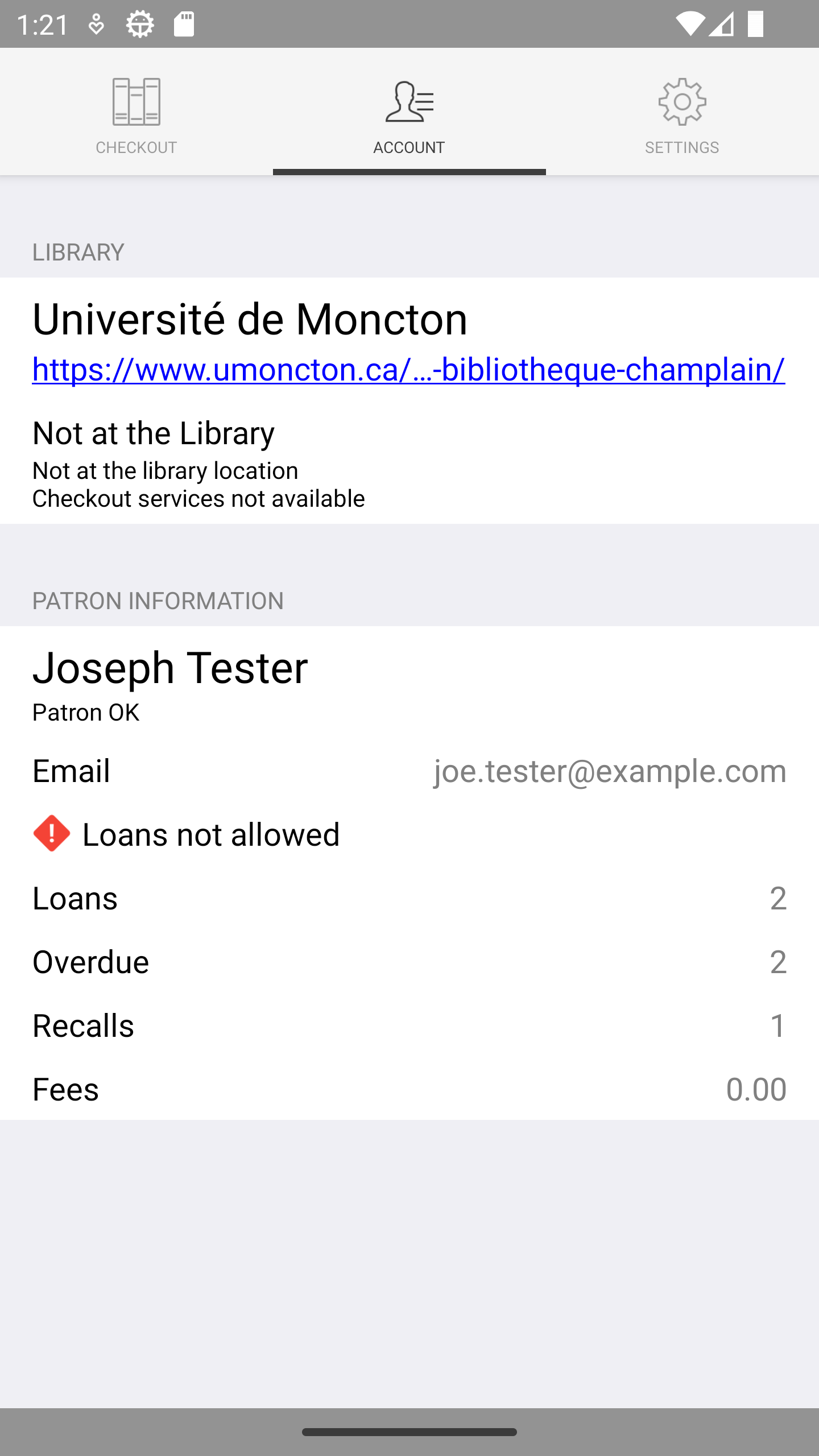}}\\
\hline
UI Generation with DMs
& 
\fbox{\includegraphics[height=3.5cm,keepaspectratio]{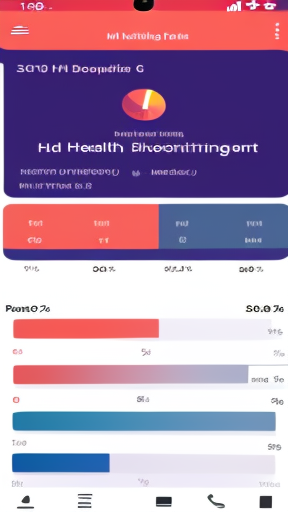}} & 
\fbox{\includegraphics[height=3.5cm,keepaspectratio]{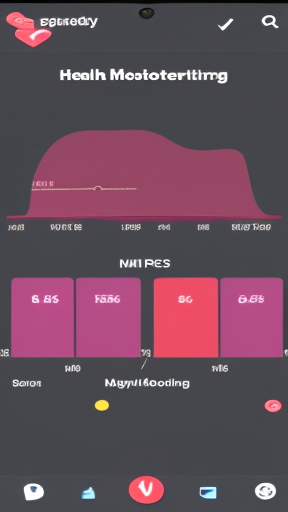}} & 
\fbox{\includegraphics[height=3.5cm,keepaspectratio]{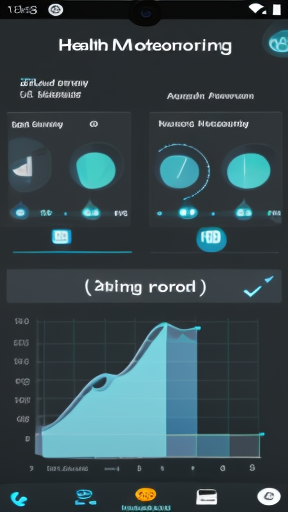}} & 
\fbox{\includegraphics[height=3.5cm,keepaspectratio]{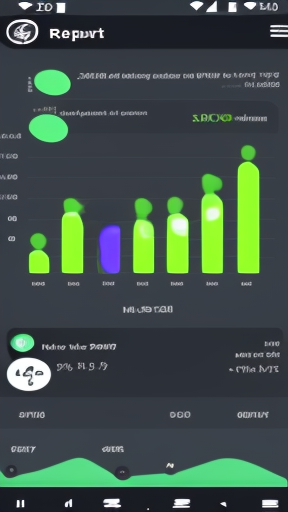}} &
\fbox{\includegraphics[height=3.5cm,keepaspectratio]{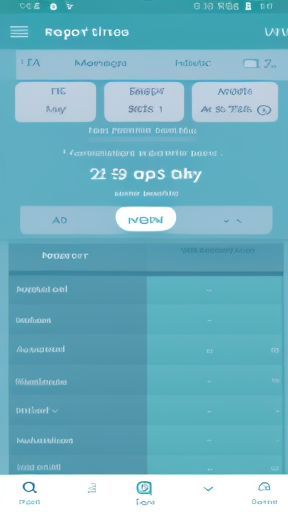}} &
\fbox{\includegraphics[height=3.5cm,keepaspectratio]{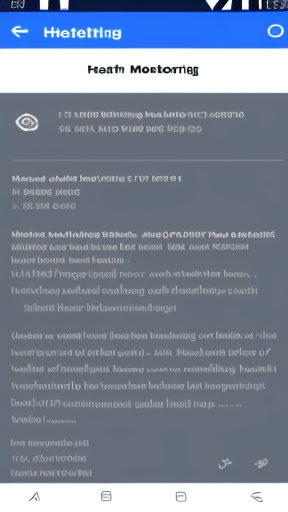}}\\
\hline[1pt]
\end{tblr}
\label{tab:samples}
\end{table*}

\section{UI Generation with LLMs}

Large Language Models (LLMs), such as GPT, represent a category of AI models designed for interpreting, generating, and manipulating human language.
These models, characterised by their ability to generate human-like text based on extensive training data, have shown remarkable proficiency in a wide range of applications, including automated content creation, customer service, and code generation \cite{Zhao:SurveyLargeLanguage:2023}. 
LLMs are based on deep learning techniques, particularly neural networks with many layers.

Given an app page description, it is possible to automatically generate UIs, e.g.~HTML code, by prompting large language models.
In a recent study, Fend et al.~fine-tuned GPT-3.5 model for UI generation \cite{Feng:DesigningLanguageWireframing:2023}.
Our preliminary assessment indicates that GPT-4o can achieve excellent results even without fine-tuning. 
We thus focus on discussing GPT-4o in the following.

The process of UI generation using LLM  includes three steps as illustrated on Table \ref{tab:comparison}:

\subsubsection{Description Refinement:}
In this step, the LLM refines the high-level feature or UI description into a list of detailed UI sections.
For example, ``health monitoring report'' as input will be refined to ``header section'', ``profile section'', ``summary section'', ``charts section''.
Each section includes a description and a corresponding code example.
This refinement ensures that the generated HTML is detailed and structured.

\subsubsection{UI Code Generation:}
In this step, the LLM uses detailed UI sections as input to generate the corresponding HTML code. 
Given the complexity of code generation, this necessitates an advanced LLM to increase the likelihood of producing high-quality HTML output.
Our preliminary assessment indicates that GPT-4o significantly outperforms GPT-3.5 in this regard.

\subsubsection{UI Code Adjustment:}
The generated HTML code might lack essential UI elements, such as a footer, or have incorrect image sizes. 
This step involves adapting the code with additional prompts to address such issues.
For example, by asking GPT-4o to ``add the footer'' to the generated HTML code, it will include a ``footer section'' as illustrated on Table \ref{tab:comparison}.

\subsubsection{Pros}
\begin{itemize}
\item \textbf{Reusable code:} 
The output of this approach is UI code as HTML, which can be partially or fully reused in subsequent development. 
The generated code can be easily modified using additional prompts. 

\item \textbf{Low hardware requirements with cloud LLMs:}
The hardware requirements for deploying this approach are minimal when using a cloud-based LLM such as GPT-4o.
\end{itemize}

\subsubsection{Cons}
\begin{itemize}
\item \textbf{Aesthetic issues:} 
Occasionally, the generated HTML exhibits alignment issues, negatively impacting its aesthetics. 
Sometimes, the generated HTML might lack CSS styling.

\item \textbf{No images:}
Although HTML code generated by the LLMs include icons provided by Material 3 library\footnote{\url{https://m3.material.io/styles/icons/overview}}, LLMs are incapable of generating images. 
This may constrain the diversity of UIs created with this approach.

\item \textbf{High latency:}
In our preliminary assessment, it takes more than 30s to generate a UI with GPT-4o.
This can be mitigated by parallelising generations.
However, each individual generation still requires a significant time.

\item \textbf{Privacy issues with cloud LLMs:} 
Utilizing cloud LLMs like GPT-4o or Claude-3\footnote{\url{https://www.anthropic.com/news/claude-3-family}} raises data privacy concerns, as it necessitates uploading user data to the cloud. 

\item \textbf{High hardware requirements with local LLMs:} 
Deploying a local LLM, such as Llama 3\footnote{\url{https://llama.meta.com/llama3/}}, requires expensive hardware, making it a less viable option for many users.
\end{itemize}

\subsubsection{Keep in mind}
\begin{itemize}
\item \textbf{Tuning the temperature of LLMs:}
The temperature setting of an LLM (ranging from 0 to 2) adjusts the predictability of its responses, directly affecting the diversity of the generated UIs. 
A higher temperature (greater than 1.0) leads to more diverse outputs but increases the likelihood of errors, while a lower temperature (less than 1.0) makes the output more deterministic, reducing variability.
\end{itemize}

\begin{table*}[]
\setlength{\fboxrule}{0pt}

\centering
\caption{Comparison of three AI-based approaches for recommending UI designs.}
\begin{tabular}{p{5cm} | p{5cm} | p{5cm}}
\hline
\textbf{GUI Generation with LLMs} & \textbf{GUI Retrieval with VLMs} & \textbf{GUI Generation with DMs} \\
\hline
\fbox{\includegraphics[width=5cm]{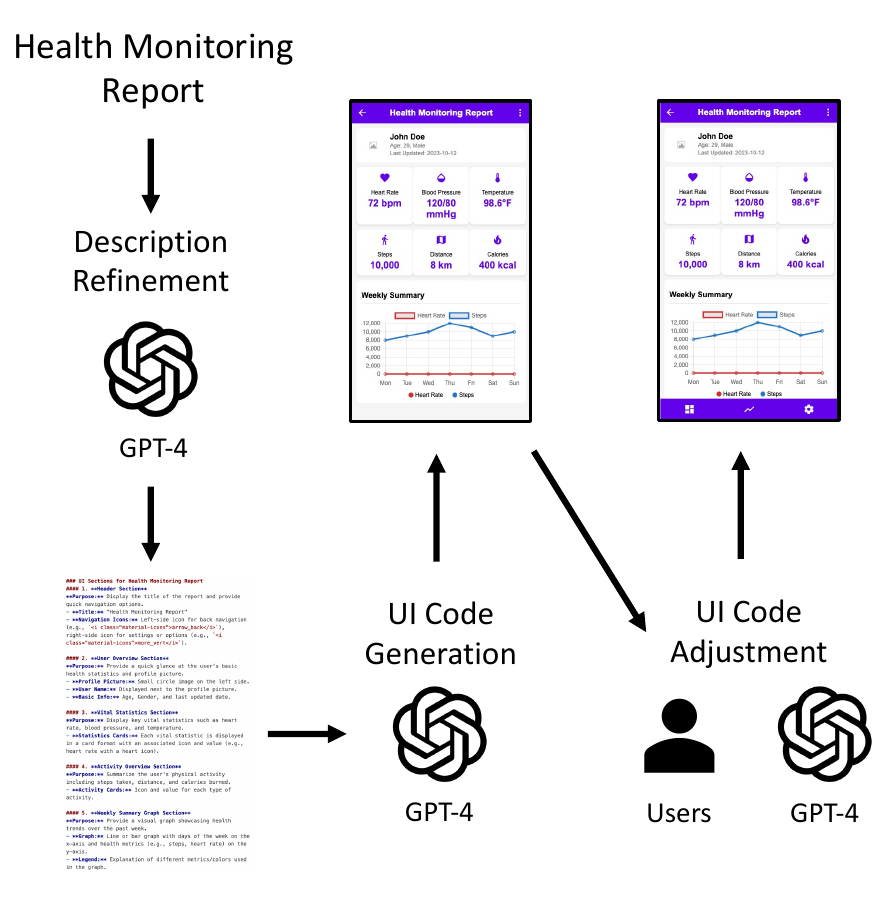}} &
\fbox{\includegraphics[width=5cm]{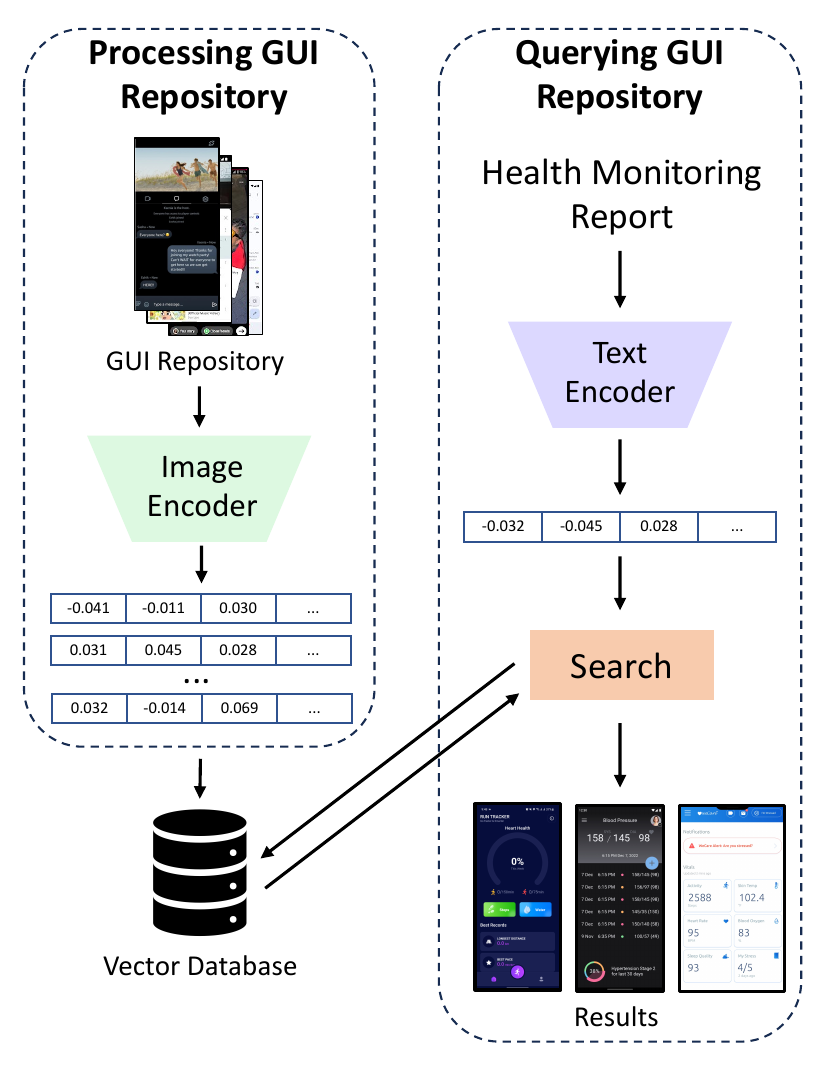}} &
\fbox{\includegraphics[width=5cm]{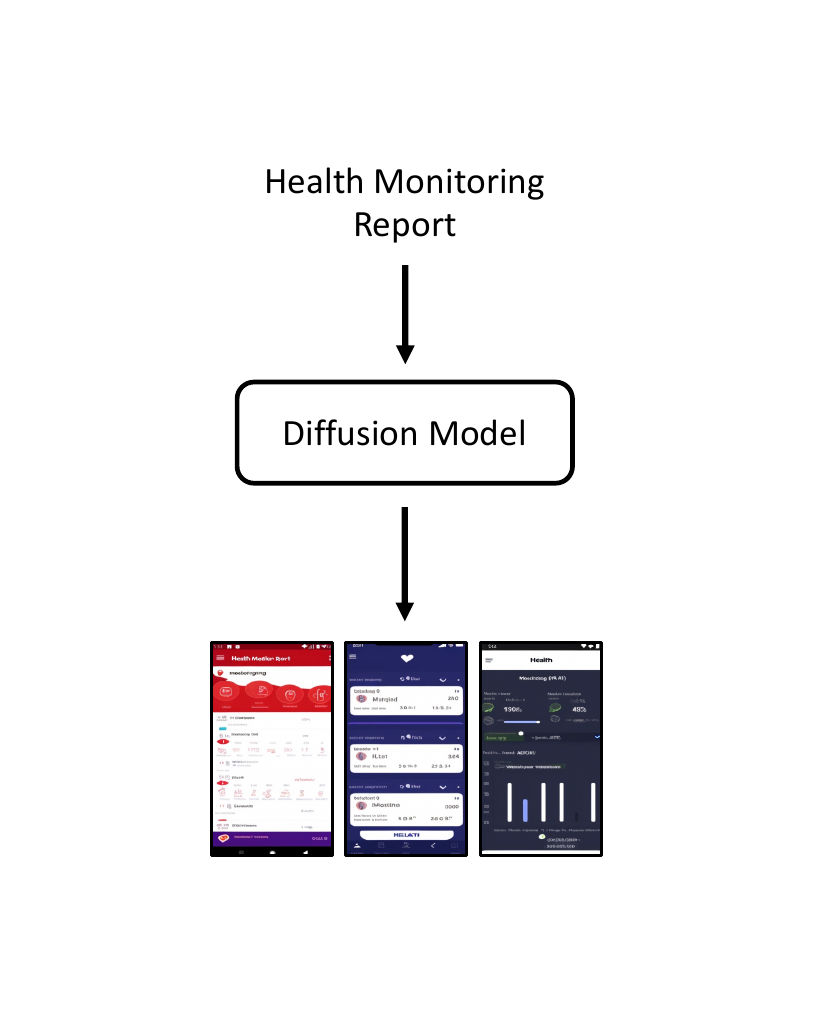}} \\

\hline
\multicolumn{3}{c}{\textbf{Diversity}} \\
\hline
Medium, no image & Limited due to number of existing apps & High diversity in layout and style \\

\hline
\multicolumn{3}{c}{\textbf{Quality}} \\
\hline
Generally good, may include alignment and style issue & High design quality and aesthetics as created by experts & Generally good, may contain unreadable text \\

\hline
\multicolumn{3}{c}{\textbf{Relevance}} \\
\hline
High & Medium, sometime irrelevant GUIs retrieved & High \\

\hline
\multicolumn{3}{c}{\textbf{Reusability}} \\
\hline
Reusable HTML code & Image format, hard to reuse; copyright restrictions & Image format, hard to reuse; aesthetics issue \\

\hline
\multicolumn{3}{c}{\textbf{Latency}} \\
\hline
\makecell[l]{Cloud LLMs: $>$30s with GPT-4o \\ Local LLMs: depends on hardware} & 
Depends on hardware; $<$1s with NVIDIA T4 &
\makecell[l]{Cloud DMs: $\approx$20s with DALL·E 3 \\
Local DMs: depends on hardware \\ $\approx$3s with NVIDIA T4 \& batch size of 1} \\

\hline
\multicolumn{3}{c}{\textbf{Hardware Requirements for Local Deployment}} \\
\hline
\makecell[l]{Require powerful GPU \\ (NVIDIA A100)} & 
\makecell[l]{Require GPU (2GB VRAM) \\ and disk space} &
\makecell[l]{Require GPU (at least \\ 4GB VRAM)} \\

\hline
\end{tabular}
\label{tab:comparison}
\end{table*}

\section{UI Retrieval with VLMs}
Instead of generating new UIs, an alternative approach is to search a UI database (e.g. App stores like Google Play or Apple App Store) using textual queries.
However, current searches operate at the app level (i.e.~the text describing the app) rather than the UI level.
To tackle this problem, Kolthoff et al.~proposed RaWi \cite{Kolthoff:DatadrivenPrototypingNaturallanguagebased:2023} a UI retrieval approach using text embedding from metadata available about the UI.
Our recent work demonstrates that VLMs clearly surpasses text embedding model in text-to-UI retrieval \cite{Wei:GUingMobileGUI:2024}.

Vision-Language Models, such as CLIP \cite{Radford:LearningTransferableVisual:2021}, are multimodal models, capable of learning both images and text (i.e.~multiple modalities). 
These models are pre-trained on large-scale \textit{image-caption} datasets using contrastive learning techniques. 
VLMs have the capability to convert images and text into a shared multimodal embedding space, ensuring that semantically similar images and texts are closely aligned. 
By computing the similarity between images and textual queries, VLMs can be effectively used in various tasks, such as zero-shot image classification (e.g., iOS vs.~Android Design) or text-to-image retrieval. 
The latter requires the following four steps.

\subsubsection{VLM Fine-tuning:}
The vanilla CLIP model \cite{Radford:LearningTransferableVisual:2021}, originally trained on general image-caption datasets that including dogs, buildings, and other common subjects, may not perform well in text-to-UI retrieval task. 
To achieve better performance, the VLM should be fine-tuned with a UI-caption dataset, such as SCapRepo \cite{Wei:GUingMobileGUI:2024}. 
Instead of fine-tuning your own VLM model, we have made our GUIClip model publicly available\footnote{\url{https://huggingface.co/Jl-wei/guiclip-vit-base-patch32}}.

\subsubsection{UI Repository Preparation:}
UI retrieval is performing a search on an existing UI dataset. 
Thus, the first step is to prepare a UI repository.
There are existing large UI datasets, like Rico \cite{Deka:RicoMobileApp:2017} and SCapRepo \cite{Wei:GUingMobileGUI:2024}.
Besides, designers can create their own repository, e.g.~though automatic UI exploration \cite{Chen:AutomaticallyDistillingStoryboard:2023}.

\subsubsection{UI Image Processing:}
UI retrieval using VLM is achieved by calculating the similarity between the embeddings of the query and the UIs.
Thus, pre-computing the embeddings of the images in the repository is essential.
As shown on Table \ref{tab:comparison}, all images in the repository are converted into embeddings with the VLM's image encoder. 
The embeddings are stored in a vector database as Faiss\footnote{\url{https://github.com/facebookresearch/faiss}} for subsequent searches.

\subsubsection{UI Image Querying:}
While using the UI search engine, the user query is converted into a text embedding using the text encoder of the VLM. 
The resulting text embedding is compared with image embeddings stored in the vector database. 
The top-k most similar images are selected as output.

\subsubsection{Pros}
\begin{itemize}
\item \textbf{High quality UI images:}
The UI images originates from existing apps. 
Thus, their quality is usually much better than those generated by an AI model.

\item \textbf{Low latency:}
The UI retrieval takes less than one second to get the results.
It is significantly faster than UI generation and might thus be useful particularly for quick exploration and iterations.

\item \textbf{Link to source app:}
Each retrieved UI image is linked to its source app, which enables developers to explore implementation details and to gather user feedback on the design.
\end{itemize}

\subsubsection{Cons}
\begin{itemize}
\item \textbf{Limited diversity:}
Unlike the generation which leverages randomness to produce unlimited number of UIs, the retrieval approach recommend UIs from existing apps, thereby limiting diversity.
It is thus important to explore loosely related apps and designs by variating the queries.

\item \textbf{Irrelevant results:}
Some retrieved UIs may be entirely irrelevant to the given query.

\item \textbf{Complexity of deployment:}
Deploying an own UI search engine necessitates several hundred gigabytes of disk space to store a large UI repository.
Additionally, maintaining a vector database is necessary for its operation.
\end{itemize}

\subsubsection{Keep in mind}
\begin{itemize}
\item \textbf{Copyright}
The retrieved UIs are likely not directly reusable in commercial products due to copyright and license restrictions.
\end{itemize}

\section{UI Generation with DMs}
Diffusion models (DMs), such as Stable Diffusion \cite{Rombach:HighResolutionImageSynthesis:2022}, are widely used in text-to-image generation.
These models operate by iteratively refining an initial random noise image through a series of denoising steps guided by the input text. 
The core principle behind diffusion models is to reverse the diffusion process, whereby an image is progressively degraded into noise. 
By training on a large images-caption dataset, the model learns to generate coherent and visually appealing images that match the given textual input.

Using text-to-image models, it is possible to generate UI images from app  descriptions. 
Wei et al.~recently introduced UI-Diffuser \cite{Wei:BoostingGUIPrototyping:2023}, a UI image generator built on Stable Diffusion \cite{Rombach:HighResolutionImageSynthesis:2022}. 
UI-Diffuser necessitates both a page description and the image layout information to achieve reasonable performance. 
With this paper, we release UI-Diffuser-V2\footnote{\url{https://huggingface.co/Jl-wei/ui-diffuser-v2}}, which generates relevant UI images using only the page description of the apps.

The UI generation can be achieved via cloud-based image generator, such as Midjourney\footnote{\url{https://www.midjourney.com/home}} and DALL·E 3\footnote{\url{https://openai.com/index/dall-e-3/}}.
However, these generators are not specifically trained for creating mobile UIs, often resulting in skewed UI designs and a higher latency.
For optimal performance, fine-tuning a diffusion model with a UI-caption dataset is recommended.

We fine-tuned the Stable Diffusion model with the recently available SCapRepo dataset \cite{Wei:GUingMobileGUI:2024}.
SCapRepo is a large dataset with 135k UI-caption pairs.
It was created by automatically performing classification, cropping, and caption extraction from the app introduction images of Google Play.
We trained the Stable Diffusion model for five epochs on a machine equipped with an NVIDIA T4 GPU, resulting in the UI-Diffuser-V2 model.

The model usage is straightforward: by inputting a page description, it generates corresponding UI images. 
It is also capable of producing multiple images in one generation, though this requires additional VRAM. 
Our preliminary assessment indicates that an NVIDIA T4 can generate over 20 images simultaneously.

\subsubsection{Pros}
\begin{itemize}
\item \textbf{High diversity:}
This approach can generate highly diverse UI images with varying layouts, providing an excellent source of inspiration.
\end{itemize}

\subsubsection{Cons}
\begin{itemize}
\item \textbf{Aesthetic issues:}
There will likely be graphical issues on the UI generated with DMs, such as unreadable text and misaligned components.
In some instances, the DM generates significantly flawed images.

\item \textbf{Limited reusability:}
The aesthetic issues coupled with the binary images hinder the immediate reusability of UIs created with DMs.
\end{itemize}

\subsubsection{Keep in mind}
\begin{itemize}
\item \textbf{Tuning the batch size:}
Batch size refers to the number of images obtained in a single generation.
Using the NVIDIA T4, it is capable of generating a UI within 3 seconds when the batch size is set to 1. 
A batch size of 20 increases the generation time to approximately 30 seconds.
\end{itemize}

\section{Limitations and Future Directions}

The presented AI approaches trigger designers' creativity by suggesting alternative designs for a certain text input.
Our preliminary evaluation indicates that the AI-suggested UIs is relevant to the input \cite{Wei:GUingMobileGUI:2024}. 
However, none of the approaches is capable of producing ready-to-use UIs. 
This might change with future multimodal models trained also on UI source code.
Still, the discussed approaches represent a valuable design tool.
Particularly for rapid ideation and prototyping, AI can swiftly produce designs based on minimal inputs, allowing stakeholders to explore diverse design ideas in a short time. 
Unconventional solutions or requirements that have not yet been considered (such as accessibility features) is likely to spark creativity.

The performance of the proposed approaches heavily rely on the underlying models and the training data. 
GPT-4o was trained on a vast corpus of internet data.
The GUIClip and UI-Diffuser-V2 models were trained on 135k screenshot-caption pairs from the SCapRepo dataset.
Although these models encode extensive knowledge from various domains, they may also introduce potential biases inherent in the training data. 
For instance, an AI model trained primarily on skeuomorphic designs may struggle to generate flat designs, which may raise fairness or reliability concerns. 
Thus, human critical thinking remains key to evaluate and eventually adjust the AI UI suggestions.

Moreover, as our previous work shows \cite{Wei:GUingMobileGUI:2024}, fine-tuning general-purpose models with custom datasets may be necessary to achieve optimal results.
At the same time, using web data for model training may raise legal and ethical issues, particularly around the licensing and copyright.
A related issue to be considered in practice is whether to use local or cloud-based AI models. 
Deploying local models requires expensive hardware, but users can fine-tune these models with custom datasets, e.g.~from previous corporate-specific designs. 
In contrast, using cloud-based models do not require such hardware but are less customizable and may pose privacy concerns.
In general, more work is needed to create diverse, rich, open datasets for recommending and evaluating UI designs addressing various usage contexts and various user needs \cite{Pourasad:ICSE:2024}.

Another important future direction is to evaluate and possibly quantify the quality of generated designs and the effectiveness of the AI-inspired design processes. 
For instance, it is unclear how much impact the input text has on the generated design.
In our preliminary tests, inputs ranged from brief phrases to sentences containing dozens of words. 
The generated outputs consistently maintained relevance to the input. 
While this indicates the flexibility of the models to deal with even not-well articulated input, an actual rigorous empirical investigation remains an open research question. 
A dedicated evaluation for various design quality measures as usability, inclusiveness, accessibility, or aesthetics is also missing.
Moreover, the overall AI-inspired approach should be effective for designers and boost the overall productivity. 
This is also yet to be explored, e.g.~depending on the designers experience or requirements severity.
Thereby a seamless integration of the AI-based approaches into the design and engineering workflows remains a critical goal. 
One possible workflow for such integration was presented in Figure \ref{fig:aid-process}.

Finally, while these AI approaches can enhance creativity by serving as sources of inspiration, the over-dependence on AI could pose a risk to human creativity \cite{HBR:2024}. 
Human creativity could get constrained by the patterns and outputs of AI models, which are ultimately limited by their training data and cannot innovate beyond what they have been exposed to. 
Thus, maintaining a balance between AI assistance and human ingenuity is essential---ensuring that AI serves as a tool for augmentation rather than replacement \cite{HBR:2024, Pourasad:ICSE:2024}.

\section{Conclusion}
Foundation models bear great potential to revolutionise software development by generating various artefacts including source code, documentation, or test cases. 
One particular artefact that usually requires creativity and multiple ideation iterations is the design of the UI. 
We discussed how to use large language models, diffusion models, and vision-language models to retrieve and generate relevant, diverse, and inspiring designs. 

Each approach has pros and cons. 
None of them is perfect. 
In practice, app teams have to carefully try and eventually combine multiple approaches---both during individual as well as group brainstorming sessions as Figure \ref{fig:aid-process} shows. 
Our preliminary assessment is based on related work and a running example shown on Table \ref{tab:samples}. 
More research is needed to gain evidence on how to combine the approaches as well as important factors to consider such as team size, domain, novelty of the features, or designers skills. 
While AI can significantly enhance creativity by serving as sources of inspiration, it cannot replace human creativity and experience. 
Human involvement (i.e.~designer, developer, and other stakeholders) remains essential to the design process \cite{HBR:2024, Pourasad:ICSE:2024}.

\def\refname{References}
\printbibliography

\begin{IEEEbiography}
{Jialiang Wei} is a Ph.D. candidate in Software Engineering at the EuroMov Digital Health in Motion (DHM) Research Unit, IMT Mines Alès.
His research focuses on requirements engineering, graphical user interface, and app store mining. 
He holds a master’s degree in computer science from the University of Technology of Belfort-Montbéliard. 
Contact him at jialiang.wei@mines-ales.fr.
\end{IEEEbiography}

\begin{IEEEbiography}
{Anne-Lise Courbis} is an assistant professor at the EuroMov Digital Health in Motion (DHM) Research Unit, IMT Mines Alès. Her research areas include Model-Based System Engineering and Formal Verification of System Architectures and Behaviours. More recently, she has become interested in using AI techniques to improve the Requirements Engineering stage.
Contact her at anne-lise.courbis@mines-ales.fr.
\end{IEEEbiography}

\begin{IEEEbiography}
{Thomas Lambolais} is an assistant professor at the EuroMov Digital Health in Motion (DHM) Research Unit, IMT Mines Alès. His research focuses on the early stages of software development, specifically Requirements Engineering and the application of formal methods to describe the behavioural properties of reactive systems. He has explored the incremental development of formal behaviours, emphasising the preservation of liveness properties. Currently, he is working on approaches for eliciting requirements and defining the knowledge and assumptions of the system environment.
Contact him at thomas.lambolais@mines-ales.fr.
\end{IEEEbiography}

\begin{IEEEbiography}
{Gérard Dray} is a professor at the EuroMov Digital Health in Motion (DHM) Research Unit, IMT Mines Alès. 
In his research, he develops information processing methods to computerize multimodal cumulative knowledge in order to facilitate human action, making it more reliable and more efficient. 
Within the framework of EuroMov DHM, he is in charge of the “factory” transverse axis, which aims to improve the reproducibility of clinical results, accelerate translational research, and technology transfer by producing standardised and documented approaches.
Contact him at gerard.dray@mines-ales.fr.
\end{IEEEbiography}

\begin{IEEEbiography}
{Walid Maalej} is a professor of informatics at the University of Hamburg, Germany, where he leads the Applied Software Technology group.
His research interests include User Feedback, Empirical Software Design and AI Engineering. 
Maalej received a Ph.D.~from the Technical University of Munich and is currently the steering committee chair of the IEEE Requirements Engineering conference. 
Contact him at walid.maalej@uni-hamburg.de.
\end{IEEEbiography}

\end{document}